\documentclass[12pt]{article}

\usepackage{setspace} 

\usepackage{afterpage}
\usepackage{algorithm}
\usepackage{algpseudocode}
\usepackage{amsbsy, amsfonts, amsgen, amsopn, amssymb, amstext, amsxtra, bezier, color, verbatim, pictexwd, supertabular, url, dsfont,leftidx, mathrsfs,setspace}
\usepackage{array}
\usepackage{amsmath}
\usepackage{bbm}
\usepackage{booktabs} 
\usepackage{caption}
\usepackage{csquotes}
\usepackage{enumerate}
\usepackage{fancyhdr}
\usepackage{graphicx}
\usepackage{hyperref}
\usepackage{lipsum}
\usepackage{longtable} 
\usepackage{mathtools}
\usepackage{multirow}
\usepackage{setspace}
\usepackage{subcaption}
\usepackage{tabularx}
\usepackage{tikz}
\usepackage[a4paper,margin=1in]{geometry} 
\usepackage{natbib}
 \bibpunct[, ]{(}{)}{,}{a}{}{,}%
 \def\bibfont{\small}%

\usepackage[compress]{cleveref}
\usepackage[title]{appendix}

\usepackage{setspace}        
\usepackage{amsthm}          
\usepackage{amssymb}         
\usepackage{tikz}             
\usetikzlibrary{positioning}

\usetikzlibrary{positioning}
\Crefname{subfigure}{Subfigure}{Subfigures}

\crefrangelabelformat{subfigure}{#3#1#4--#5\crefstripprefix{#1}{#2}#6}

\usepackage[normalem]{ulem}
\usepackage[title]{appendix}

\begin{document}
\pagenumbering{gobble}
\begin{titlepage}
    \centering
    \vspace*{1cm}

    \Huge
    \textbf{Corporate Fundamentals and Stock Price Co-Movement}

    \vspace{1.5cm}

    \small
    Lyuhong Wang

    \vspace{0.5cm}

    \textnormal{Trinity College Dublin}

    \vspace{0.5cm}

    \textnormal{E-mail: lywang@tcd.ie}

    \vspace{1cm} 

    \small
    Jiawei Jiang

    \vspace{0.5cm}

    \textnormal{Shenzhen HTI Group Co.,Ltd.}

    \vspace{0.5cm}

    \textnormal{E-mail: j\_jw@hotmail.com}

    \vspace{1cm} 

    \small
    Yang Zhao

    \vspace{0.5cm}

    \textnormal{Suzhou University of Science and Technology}

    \vspace{0.5cm}

    \textnormal{E-mail: zhaoyang@mail.usts.edu.cn}

    \vfill

    \vspace{0.8cm}

\end{titlepage}

\onehalfspacing 

\section*{Abstract}
We introduce an innovative framework that leverages advanced big data techniques to analyze dynamic co-movement between stocks and their underlying fundamentals using high-frequency stock market data. Our method identifies leading co-movement stocks through four distinct regression models: Forecast Error Variance Decomposition, transaction volume-normalized FEVD, Granger causality test frequency, and Granger causality test days. Validated using Chinese banking sector stocks, our framework uncovers complex relationships between stock price co-movements and fundamental characteristics, demonstrating its robustness and wide applicability across various sectors and markets. This approach not only enhances our understanding of market dynamics but also provides actionable insights for investors and policymakers, helping to mitigate broader market volatilities and improve financial stability. Our model indicates that banks' influence on their peers is significantly affected by their wealth management business, interbank activities, equity multiplier, non-performing loans, regulatory requirements, and reserve requirement ratios. This aids in mitigating the impact of broader market volatilities and provides deep insights into the unique influence of banks within the financial ecosystem.

\section*{Keywords}
Corporate Fundamentals, Stock Price Co-movement, Stock Price Synchronization, Stock Price Homogenization

\section{Introduction}\label{sec:Intro}
The co-movement of stock prices in the financial market refers to the correlation and linkage between the prices of different financial assets \citep{engleDynamicConditionalCorrelation2002}. This co-movement reflects the influence of factors such as information transmission, investor behavior, and market structure. In this paper, we focus on studying the co-movement of stock prices within the same industry. Studying price co-movement offers insights into the underlying mechanisms of financial markets, providing investors with more precise market forecasts and risk management strategies.

Contrary to the prevailing focus in existing literature on stock price co-movement analysis using daily or lower frequency data, our study addresses a notable gap by proposing a novel framework based on high-frequency data. This framework systematically and rigorously investigates the relationship between stock price co-movement and fundamental characteristics within the same industry. Previous research often neglects the impact of concurrent market index fluctuations on industry-specific co-movement. Additionally, there is a scarcity of studies considering the role of industry-specific fundamental data in analyzing stock price co-movement. The approach of our framework enables an in-depth analysis of factors influencing co-movement within an entire industry. In our study, we focus on the banking sector to validate the effectiveness of our innovative framework and analytical methods. The distinct market dynamics and the extensive availability of relevant data within the banking sector provide a robust testing ground for our approach. By analyzing the nuances of stock price co-movements in this context, particularly within high-frequency trading environments, we demonstrate the applicability and strength of our model in uncovering industry-specific insights.

Understanding stock price co-movement is particularly important for the banking sector due to its pivotal role in the financial system. Banks serve as crucial intermediaries in the economy and act as barometers of financial stability. Their unique market dynamics, including interactions with monetary policy, regulatory environments, and macroeconomic conditions, make them an ideal subject for studying price co-movement.

The selection of the banking industry as the research subject is significant for several reasons. Firstly, as a core component of the economic system, the banking industry critically influences the stability and operational efficiency of the entire financial system. Secondly, the banking industry’s distinctive business model and risk characteristics, as noted by \citep{greenwoodVulnerableBanks2015}, make it a unique sector for study. Thirdly, the highly homogenized business model among Chinese banks \citep{Berger_Hasan_Zhou_2010} offers a conducive environment for identifying common patterns in stock price movements. The uniformity in core business practices, such as loan issuance, wealth management, and interbank transactions, reduces the variables that could potentially influence stock price behavior, facilitating a more streamlined and focused analysis. This allows for a more accurate identification of systemic factors and trends that influence not just individual banks but the industry as a whole \citep{Demirgüç-Kunt_Pedraza_Ruiz-Ortega_2021}.

Studies have highlighted various dimensions of this co-movement. For instance, \citep{Wang_Zhang_Li_Chen_Wei_2020}  demonstrated that sharp declines in Chinese bank stock prices not only trigger systemic financial risks in China but also destabilize global financial stability. Similarly,  \citep{Acharya_Engle_Jager_Steffen_2024} showed that balance sheet liquidity risk was a major driver of U.S. bank stock returns during the initial phase of the COVID-19 pandemic, underscoring the sensitivity of bank stocks to liquidity conditions and the broader implications for financial markets. The collapse of Silicon Valley Bank had systemic repercussions not only within the banking industry but also across global capital markets, as evidenced by \citep{Aharon_Ali_Naved_2023, Jiang_Matvos_Piskorski_Seru_2024}.

Furthermore, \citep{Borri_Giorgio_2022} provided evidence of a strong linkage between sovereign default risk and the systemic risk of banks, emphasizing the dual role banks play in both reflecting and propagating financial instability. These studies collectively underscore the importance of examining bank stock price co-movements to gain insights into the broader financial ecosystem. By understanding how bank fundamentals interact with stock prices, investors and policymakers can better anticipate and mitigate systemic risks. This is particularly relevant in a high-frequency trading environment where rapid information dissemination and market reactions can amplify the effects of shocks.

However, while there is a wealth of literature on stock price co-movement and the impact of bank fundamentals on stock returns, the exploration of the determinants of this co-movement, especially the role of bank fundamentals, remains relatively limited. By focusing on these determinants, our study provides valuable insights that can aid in mitigating the impact of broader market volatilities and enhancing risk management strategies within the banking sector. This research is not only academically significant but also holds practical relevance for investors, policymakers, and financial institutions aiming to understand and navigate the complexities of market behavior. 

In terms of bank-specific factors, a significant body of literature has examined the effects of earnings quality, transparency, and risk-taking behavior on bank stock returns \citep{demirguc-kuntBankActivityFunding2010, pradhanCausalNexusEconomic2014}. While research on the correlation between bank fundamentals and stock returns is abundant and varied, certain factors like asset quality, capital adequacy, earnings quality, and liquidity stand out as notable influencers. However, their relationship with stock price co-movement has yet to be fully explored.

We aim to bridge this gap by employing methods such as the FEVD (Forecast Error Variance Decomposition) method and the Granger causality relationship \citep{grangerInvestigatingCausalRelations1969}. 

In this paper, we delve into the intricate dynamics between the co-movement of stock prices and underlying bank fundamentals, with a particular focus on Chinese listed banking institutions. By selecting the banking sector as our primary field of investigation, this study aims to elucidate not only the sector-specific behaviors and trends but also to validate the efficacy of our analytical framework.  

One of the objectives of this study is to identify the characteristics of market-leading banks, known as ``industry bellwethers." These bellwether stocks often have a high market impact and leadership position, and their price fluctuations may significantly influence other same-industry stocks. By identifying these bellwether stocks, we can gain better insights into the market structure and competitive landscape of the industry.

Another objective is to examine the fundamental characteristics of these banks. Bank fundamentals encompass indicators such as financial condition, profitability, asset quality, and risk management capability. Through an in-depth analysis of these fundamental characteristics, we can assess the banks' operational status, risk tolerance, and future development potential \citep{bergerProblemLoansCost1997}.

This study also aims to explore the relationship between fundamental characteristics and stock price co-movement. Through correlation analysis and regression analysis, we can evaluate the contribution of fundamental factors to the co-movement of bank stock prices. This will help us understand the role and significance of fundamental factors in industry price co-movement and provide investors with more comprehensive information and a decision-making basis for investors and regulators.

While the aforementioned studies provide a comprehensive understanding of co-movements \citep{aghabozorgiStockMarketComovement2014, barberisComovement2005, boxQualitativeSimilarityStock2018, chenComovementRevisited2016, kumarInvestorSentimentReturn2013, zhangComovementsStockPrices2017}, the influence of fundamental factors on stock price co-movement, particularly in banking, is an area ripe for exploration. Bank-specific factors, such as asset quality, capital adequacy, earnings quality, and liquidity, could potentially influence stock price co-movement. Asset quality, for instance, directly reflects a bank's risk level, which may synchronize the bank's stock return with others. Capital adequacy could serve as a signal for bank stability, which, in turn, affects the co-movement of stock prices. Similarly, earnings quality and liquidity could also lead to co-movement through their impact on investors' perception of bank risk and performance.

Our research introduces a novel framework for analyzing the fundamental data of stocks and their co-movement within industries. Empirically, our findings, based on this framework, suggest that investors show a marked preference for banks with lower capital costs, a diversified revenue structure, and minimal dependence on single income sources, all supported by a robust risk management system.

Interestingly, our framework also reveals that banks with significant exposure to high-risk sectors, yet equipped with solid risk control mechanisms, tend to more profoundly influence the stock trajectories of their industry counterparts. This highlights the intricate expectations of investors in the banking sector, emphasizing a balance between risk management, growth potential, and profitability. Banks adhering to overly conservative models that neglect potential profit opportunities tend to face investor wariness. These insights are particularly valuable for bank management and valuation, further demonstrating the practical utility of our framework.

Looking forward, we plan to apply our framework to analyze the relationship between stock price co-movement and fundamental data across various industries and countries. We hope this framework will offer new perspectives in asset pricing across different sectors. Additionally, we aim to explore its application in stock investment backtesting. For instance, we plan to integrate cyclical prediction models \citep{McMillan_2019,Yu_Huang_Chen_2023} with our framework to analyze whether buying industry-leading stocks at the bottom of their price cycles can yield additional profits.

\section{Literature Review}\label{sec:Review}
Co-movements in stock prices have long been a central point of focus across numerous studies due to their inherent complexity and multi-dimensional impact. These co-movements are typically attributed to a variety of factors such as the dissemination of market-wide information, contagion effects, and herding behavior among investors, according to the theoretical models \citep{aghabozorgiStockMarketComovement2014,greenPricebasedReturnComovement2009}. And the work of \citep{wahalStyleInvestingComovement2013} delved deeper into the influence of style investing on this comovement phenomenon. Their research suggests a crucial role of style investing in predicting returns, thereby pointing to its potential as a key driver of co-movements in stock prices.

An essential contribution to this discourse comes from \citep{kumarInvestorSentimentReturn2013}, who underscore the significant impact of both retail and institutional investors' trading activities on the co-movement of stock returns. Their research is particularly noteworthy for its exploration of events like stock splits and changes in headquarters, which can significantly influence the co-movement dynamics.

\citet{engleHedgingClimateChange2020} further expanded the scope of this discourse by examining the effect of climate change news on stock price volatility across various industries, with a keen focus on energy and finance sectors. In the context of rapidly developing countries like China, this is an increasingly important issue given the high energy demand. The authors devised and implemented a strategy to hedge against climate change risks dynamically. They successfully derived insights from a series of climate news, constructed through newspaper text analysis, and used a simulated portfolio approach to build climate change hedging portfolios. They set a standard for this process by leveraging ESG ratings from third-party companies to simulate their climate risk exposure. In parallel, the studies conducted by \citep{nicolaComovementMajorEnergy2016} illuminate the co-movement dynamics of key energy, agricultural, and food commodity price returns, contributing to our understanding of co-movement beyond stock prices.

Furthermore, \citet{zhaoDynamicRelationshipExchange2010}'s work provided significant insights into the dynamic relationship between exchange rates and stock prices in China through the employment of VAR (Vector Auto-regression) and Multivariate GARCH (Generalized AutoRegressive Conditional Heteroskedasticity) models. The study reveals that there is no stable long-term equilibrium relationship between exchange rates and stock prices in China, which adds valuable insights into the short-term dynamic relationship between the two. Intriguingly, Zhao's research also divulges the existence of volatility spillover effects between these two markets, implying that fluctuations in exchange rates can markedly affect stock prices, including those within the banking sector.

A broader investigation of co-movements has also been undertaken within the energy sector. \citet{zhangComovementsStockPrices2017} specifically examined co-movements among the stock prices of new energy, high-technology, and fossil fuel companies in China. Their work broadened the understanding of co-movement to encompass an inter-industry level, spotlighting the significant interconnections between the stock prices of companies operating in distinct sectors.

The preponderance of the existing literature is primarily focused on market-wide and industry-specific factors, often overlooking the elements causing co-movement within the banking sector. However, the work of \citet{greenwoodVulnerableBanks2015,pradhanCausalNexusEconomic2014} fills this void by exploring the underlying attributes that prompt price co-movements within the banking industry. Adding to this discourse, \citet{battistonDebtRankTooCentral2012} introduces a pioneering risk measurement methodology termed \textbf{DebtRank}. This novel metric is geared towards determining systemically important nodes in financial networks and can be instrumental in comprehending and predicting stock price volatility.

Taking a closer look at the Chinese landscape, \citet{yaoInvestorHerdingBehaviour2014} utilized data from the Chinese stock market to examine the conformity of investors. They scrutinized the evidence of herd behavior, spillover effects related to herd behavior, and the conditions under which herd behavior is driven by fundamental or non-fundamental factors. Their research found that the herd effect is prevalent in both A-share and B-share markets.

Moreover, research by \citet{huangNetworkPerspectiveComovement2021} leveraged complex network theory to analyze the co-movement and structural changes among individual stocks in the market. Based on the insights drawn from \citet{wahalStyleInvestingComovement2013}, these structural changes can be influenced by a range of factors, including style investing, fundamental elements, and industry-specific dynamics.

Broadening the discussion beyond the Chinese context, \citet{raifuReactionStockMarket2021a} delved into how the global surge in COVID-19 cases and resultant deaths affected Nigerian stock market returns. The introduction of health crises and related policy responses, such as lockdown measures, adds a crucial dimension to the analysis of price co-movements within the banking sector.

\citet{pastorUncertaintyGovernmentPolicy2012, pradhanCausalNexusEconomic2014} carried out an in-depth exploration of the causal relationship between economic growth, banking sector development, stock market development, and other macroeconomic variables, with a particular emphasis on ASEAN countries. These factors play a considerable role in the co-movement in stock returns, accentuating the need to consider a wider array of economic variables to perceive the dynamics of price co-movement within the banking industry. This becomes especially critical in light of the research on style investing by \citet{wahalStyleInvestingComovement2013}.

\citet{pastorUncertaintyGovernmentPolicy2012} provides crucial insights into how the uncertainty of government policies influences the volatility and co-activity of stock prices. This is of particular importance in countries like China, where government policies have a profound impact on market dynamics. They introduced a general equilibrium model that includes uncertainty in government policies and governments with both economic and non-economic motivations, shedding light on how policy changes impact stock prices. Additionally, \citet{parsleyReturnComovement2020} probed into the shared co-movements of returns across markets, highlighting the effects of macroeconomic policy stability, the frequency of economic crises, and trading activity levels. Their work extends the debate on co-movements in stock returns by factoring in these external influences.

Inspired by the work of \citet{kumarInvestorSentimentReturn2013}, this paper also evaluates how investors' trading activities can induce excessive co-movements in stock returns, a crucial consideration for comprehending price co-movements within the banking industry. Simultaneously, \citet{nicolaComovementMajorEnergy2016} pinpointed a high dynamic co-movement between energy and agricultural commodity price returns, introducing another important perspective when analyzing price co-movements within the banking sector.

Echoing this sentiment, this review underlines the importance of studying investor behavior trends to effectively understand price co-movements within the banking industry. The complexity and multi-faceted nature of price co-movements demand an equally multi-dimensional research approach.

In summary, the body of literature on stock price co-movements is extensive, covering a broad range of factors influencing these co-movements. However, the field remains ripe for additional exploration, particularly in light of the dynamic and rapidly evolving economic landscape. As our understanding of these dynamics deepens, we can better predict and mitigate potential risks, ultimately strengthening the stability and resilience of the global financial system.

\section{Methodology}\label{sec:Method}
Our methodology is structured into several key components: data requirements, data processing techniques, model selection, definition of dependent and independent variables, and cross-validation approaches.

\subsection{Data Requirements}
In our methodology analysis, we require one year of stock price data, recorded at five-minute intervals, for all stocks that exhibit co-movement characteristics, typically those within the same industry or sharing a common investment concept. Additionally, our study incorporates fundamental data, which is categorized into three main types: financial report data, regulatory data, and analyst forecast data.

\subsection{Data Processing}
Our methodology's data processing procedure begins with the adjustment of stock price data. We utilize a Post-Adjustment Factor to recalibrate the data for events such as stock splits and dividends, ensuring the accuracy of our analysis. The five-minute return data for each stock is then computed. To derive excess returns, our methodology asks calculate the difference between each stock's five-minute returns and the concurrent returns of a benchmark market index. This approach enables us to isolate the stock-specific performance from broader market movements.

Data cleaning is a critical step in our methodology. All stocks with excessive missing values (over 1,000 within a year) are excluded, as they equate to a substantial amount of trading time (approximately 20 trading days based on a 4-hour trading day). Our methodology apply a zero-imputation method for other sporadic missing return values. In cases of missing trading volume data, a ratio-based imputation method is employed, using the stock's previous day's trading volume ratio as an estimator.

Furthermore, our methodology process financial report data by removing entries with over 20\% missing values, while those with less than 20\% missing values are imputed using average values from similar-sized stocks. To normalize data points with high average values (over 100), logarithmic transformation is applied. Additionally, our methodology employ one-hot encoding for variables present in fewer than a certain threshold percentage of stocks (calculated as 4/n, with n being the total number of stocks). This encoding also extends to Boolean variables, enhancing categorical clarity.

\subsection{Model Selection and Variables}
Our model selection criteria revolve around cross-validating outcomes across at least two of the four proposed models. These models include Forecast Error Variance Decomposition (FEVD), transaction volume-normalized FEVD, Granger causality test frequency, and Granger causality test days. This strategy ensures a comprehensive and robust validation of our methodology.

In terms of variables, the dependent variable across all models is the series of 5-minute excess returns for each stock, providing a granular view of stock performance. The independent variables constitute a weighted aggregate of 5-minute excess returns from other stocks, with the weights assigned based on the transaction volumes of each stock. This design allows us to assess the influence of broader stock movements on individual stocks. To mitigate the influence of liquidity on stock price volatility, we normalize output metrics by their respective transaction volumes.

\subsection{Cross-Validation}
In our cross-validation framework, we evaluate the influence of fundamental data on stock price co-movements. The outcomes generated by the FEVD and Granger causality models serve as dependent variables, while the collected fundamental data act as independent variables. This structure aims to pinpoint fundamental data elements significantly impacting stock price synchronization.

A unique aspect of our methodology is our approach to multicollinearity detection. Rather than conducting standard preliminary tests, we progressively eliminate the independent variable with the highest conditional number during the regression process, continuing until the model's conditional number falls below 100. This method ensures a more organic assessment of data impact, free from premature constraints based on theoretical assumptions.

For a variable to be considered significant in identifying leading stocks in price co-movement, it must be validated in at least two of the four models. This criterion enhances the robustness and reliability of our findings by minimizing the risk of spurious correlations. We also employ statistical significance tests like p-values and confidence intervals to validate the variables within each model. Additionally, this cross-validation approach allows us to assess the generalizability of our models across different datasets and timeframes, thus improving the applicability of our methodology in varying market conditions.

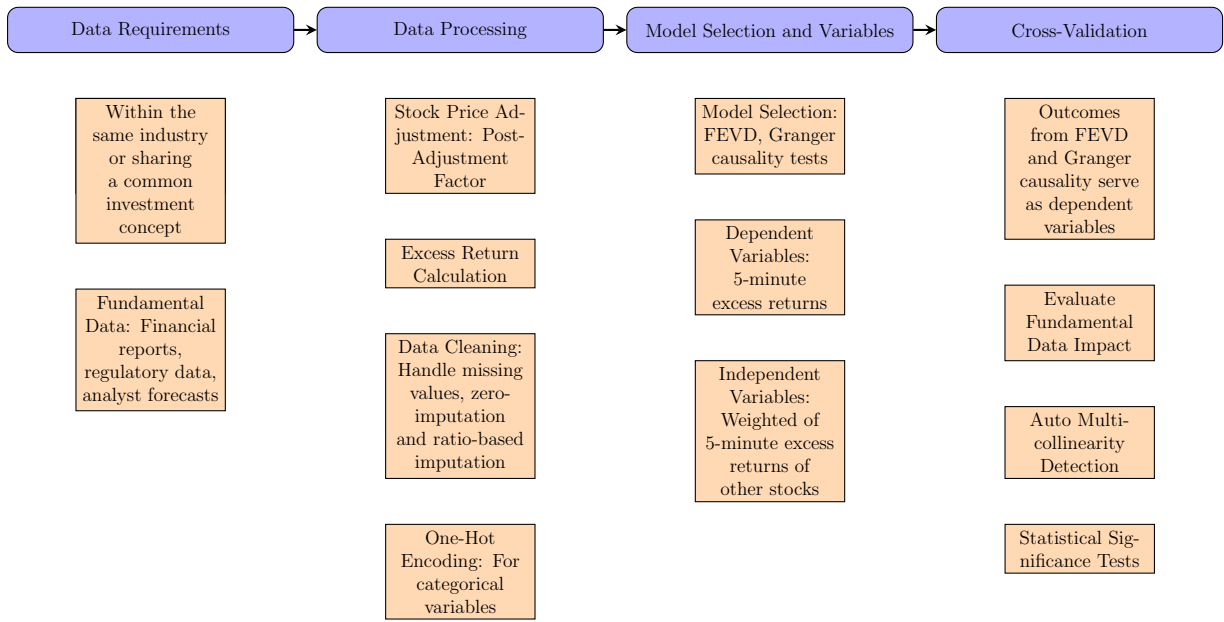
\begin{figure}[ht]
\centering
\begin{tikzpicture}[node distance=1cm and 0.5cm, auto, scale=0.6, transform shape]
\tikzstyle{section} = [rectangle, rounded corners, draw=black, fill=blue!30, text width=6cm, align=center, minimum height=1cm]
\tikzstyle{item} = [rectangle, draw=black, fill=orange!30, text width=3cm, align=center, minimum height=1cm]
\tikzstyle{arrow} = [thick,->,>=stealth]

\node[section] (datareq) {Data Requirements};
\node[item, below=of datareq] (stockdata) {Stock Price Data: One year of five-minute interval data}; 
\node[item, below=of datareq] (stockstyle) {Within the same industry or sharing a common investment concept};
\node[item, below=of stockstyle] (funddata) {Fundamental Data: Financial reports, regulatory data, analyst forecasts};

\node[section, right=of datareq] (dataproc) {Data Processing};
\node[item, below=of dataproc] (adjust) {Stock Price Adjustment: Post-Adjustment Factor};
\node[item, below=of adjust] (excessret) {Excess Return Calculation};
\node[item, below=of excessret] (dataclean) {Data Cleaning: Handle missing values, zero-imputation and ratio-based imputation};
\node[item, below=of dataclean] (onehot) {One-Hot Encoding: For categorical variables};

\node[section, right=of dataproc] (model) {Model Selection and Variables};
\node[item, below=of model] (modselect) {Model Selection: FEVD, Granger causality tests};
\node[item, below=of modselect] (devariables) {Dependent Variables: 5-minute excess returns};
\node[item, below=of devariables] (indevariables) {Independent Variables: Weighted of 5-minute excess returns of other stocks};

\node[section, right=of model] (crossval) {Cross-Validation};
\node[item, below=of crossval] (From) {Outcomes from FEVD and Granger causality serve as dependent variables};
\node[item, below=of From] (fundimpact) {Evaluate Fundamental Data Impact };
\node[item, below=of fundimpact] (multicollin) {Auto Multicollinearity Detection};
\node[item, below=of multicollin] (significance) {Statistical Significance Tests};

\draw [arrow] (datareq) -- (dataproc);
\draw [arrow] (dataproc) -- (model);
\draw [arrow] (model) -- (crossval);

\end{tikzpicture}
\caption{Flowchart of the research methodology.}
\label{fig:method-flowchart}
\end{figure}

\section{Data}
\subsection{Data Type:}
In this study, we focus on utilizing the 5-minute interval bank stock prices from the 2021 Chinese A-share market, specifically targeting the banking sector. This choice is driven by several key factors: the accessibility of data from the Chinese A-share market, the comprehensive coverage of banking stocks by securities analysts, stringent regulatory oversight, and the high degree of operational homogeneity among Chinese banks \citep{Berger_Hasan_Zhou_2010}. 

These characteristics not only facilitate our data acquisition and analysis but also enhance the robustness of our framework's testing. 

\subsection*{Post-Adjustment Method}
The Post-Adjustment method is employed to calculate the return on investment. This method presupposes an investment strategy where the investor divests all holdings a day before the ex-dividend date and reinvests the entire proceeds on the ex-dividend date at the prior day's closing price, thus not partaking in the dividend distribution. This approach ensures the full utilization of the initial investment, which remains unaffected by dividend payments or rights issues. Given the high dividend payout ratio and frequent rights issues in the banking sector, excluding these elements from the stock price analysis is essential to maintain the integrity of our research findings.

In the pursuit of designing a formula that encapsulates the outlined description, it is imperative to incorporate a multitude of critical factors. Firstly, the \textbf{Original Investment} must be accounted for, signifying the initial capital allocated in the acquisition of the stock. Secondly, paramount to our analysis is the \textbf{Rate of Return}, which stands as the principal metric of interest in our data. Thirdly, the \textbf{Ex-dividend Date} plays a crucial role; this is the specific date when dividends are dispensed to shareholders, consequentially resulting in a typical reduction in the stock price by an amount equivalent to the dividend issued. Additionally, the \textbf{Closing Price of the Previous Day} – the market value of the stock after the trading session preceding the ex-dividend date – is an essential variable in this context. It's also noteworthy that certain elements, specifically the \textbf{Dividend Payout Ratio} and \textbf{Rights Issues}, have been intentionally omitted to preserve the research's integrity and focus. This deliberate exclusion aids in maintaining the "purity" of the research by eliminating factors that might otherwise introduce unwarranted complexities or confounding variables into the analysis.

Given these considerations, the rate of return \( R \) can be calculated as follows:
\subsection*{Post-Adjustment Method Formula:}
\begin{equation}
R = \frac{N_{\text{ex-div}} \times P_{\text{ex-div}} - N_{\text{prev}} \times P_{\text{prev}}}{N_{\text{prev}} \times P_{\text{prev}}} \times 100
\end{equation}
where:
\begin{itemize}
    \item \( I \) represents the initial investment.
    \item \( P_{\text{prev}} \) denotes the closing price of the stock the day before the ex-dividend date.
    \item \( N_{\text{prev}} \) is the number of shares held the day before the ex-dividend date, calculated as \( \frac{I}{P_{\text{prev}}} \).
    \item \( P_{\text{ex-div}} \) is the closing price of the stock on the ex-dividend date.
    \item \( N_{\text{ex-div}} \) represents the number of shares repurchased on the ex-dividend date, calculated as \( \frac{I}{P_{\text{ex-div}}} \).
\end{itemize}

By doing these adjustments, our data excludes the impact of dividends and rights issues, thereby aligning with the objective of maintaining the "purity" of the research as stated.

Then, this study utilizes a series of 5-minute interval excess returns (returns of the bank stocks minus the change in the Shanghai Composite Index during the same period) for all bank stocks listed on the A-shares of the Chinese Stock Market in 2021. These returns are calculated using the price change rate adjustment method, to account for the high dividend payout ratio and probability of rights issues in banks. In addition, a total of 192 categories of bank financial report data were analyzed.

\subsection{Data description:}
The data for our study were sourced from the WIND database, public financial reports of listed banks, official documents, and reports from regulatory bodies. Additionally, we utilized advanced big data techniques, such as natural language processing for extracting insights from textual data, and distributed computing systems for handling large-scale data processing. These sophisticated methods allowed us to efficiently acquire and process the extensive dataset, ensuring comprehensive coverage and high accuracy. Analysts' insights and perspectives were also incorporated to enhance the robustness of our assumptions and conclusions. As shown in Table~\ref{tab:A1}.

\subsection{Banks name}
We employed data from banks listed on the Chinese A-share market in the year 2021. Banks with more than 1000 missing values were excluded from our study. Considering that we utilize 5-minute interval data, where each trading day consists of \(4 \times 12\) five-minute intervals, any bank missing data for more than 21 trading days was omitted. For the specific names and stock codes of the banks included in our analysis, please refer to \textbf{Table~\ref{tab:BankNames}}.

\begin{table*}[htbp]
\centering
\caption{List of Bank Names and Their Corresponding Codes}
\label{tab:BankNames}
{\small 
\begin{tabular}{|c|l|l|}
\hline
\textbf{No.} & \textbf{Code}       & \textbf{Bank Name}                          \\ \hline
1   & sh.600000  & Pudong Development Bank            \\ \hline
2   & sh.600015  & Huaxia Bank                        \\ \hline
3   & sh.600016  & Minsheng Bank                      \\ \hline
4   & sh.600036  & China Merchants Bank               \\ \hline
5   & sh.600908  & Bank of Wuxi                       \\ \hline
6   & sh.600919  & Bank of Jiangsu                    \\ \hline
7   & sh.600926  & Bank of Hangzhou                   \\ \hline
8   & sh.600928  & Bank of Xi'an                      \\ \hline
9   & sh.601009  & Bank of Nanjing                    \\ \hline
10  & sh.601077  & Yu Agricultural Commercial Bank    \\ \hline
11  & sh.601128  & Bank of Changshu                   \\ \hline
12  & sh.601166  & Societe Generale Bank              \\ \hline
13  & sh.601169  & Bank of Beijing                    \\ \hline
14  & sh.601187  & Bank of Xiamen                     \\ \hline
15  & sh.601229  & Bank of Shanghai                   \\ \hline
16  & sh.601288  & Agricultural Bank                  \\ \hline
17  & sh.601328  & Bank of Communications             \\ \hline
18  & sh.601398  & ICBC                               \\ \hline
19  & sh.601528  & Ruifeng Bank                       \\ \hline
20  & sh.601577  & Bank of Changsha                   \\ \hline
21  & sh.601658  & Postal Reserve Bank                \\ \hline
22  & sh.601665  & Qilu Bank                          \\ \hline
23  & sh.601818  & Everbright Bank                    \\ \hline
24  & sh.601825  & Shanghai Agricultural and Commercial Bank \\ \hline
25  & sh.601838  & Bank of Chengdu                    \\ \hline
26  & sh.601860  & Zijin Bank                         \\ \hline
27  & sh.601916  & Zheshang Bank                      \\ \hline
28  & sh.601939  & Construction Bank                  \\ \hline
29  & sh.601963  & Bank of Chongqing                  \\ \hline
30  & sh.601988  & Bank of China                      \\ \hline
31  & sh.601997  & Guiyang Bank                       \\ \hline
32  & sh.601998  & China CITIC Bank                   \\ \hline
33  & sh.603323  & Sunon Bank                         \\ \hline
34  & sz.000001  & Ping An Bank                       \\ \hline
35  & sz.002142  & Bank of Ningbo                     \\ \hline
36  & sz.002807  & Jiangyin Bank                      \\ \hline
37  & sz.002839  & Zhangjiagang Bank                  \\ \hline
38  & sz.002936  & Bank of Zhengzhou                  \\ \hline
39  & sz.002948  & Qingdao Bank                       \\ \hline
40  & sz.002958  & Qingnong Commercial Bank           \\ \hline
41  & sz.002966  & Bank of Suzhou                     \\ \hline
\end{tabular}
}
\end{table*}

\subsection{Data Processing}
All data passed the Augmented Dickey-Fuller (ADF) and Variance Inflation Factor (VIF) tests.

The excess return of each bank was treated as the dependent variable, while the trading volume weighted return of other banks was the independent variable.

\section{Experimental Design}
By employing FEVD, we can quantitatively assess the proportion of movements in stock price that can be attributed to shocks in its own fundamentals versus those in other or external factors. This decomposition provides deeper insights into the dynamic interactions and relative importance of various influences on stock price movements, which is critical for understanding the systemic risk and interconnectedness within the banking sector. FEVD has been widely used to study spillover effects between different variables. For example, \citep{Hanif_Hernandez_Mensi_Kang_Uddin_Yoon_2021} employed FEVD to study the dependence between clean/renewable energy sector stocks and European emission allowance prices. Additionally, \citep{Wang_Zhang_Li_Chen_Wei_2020} used FEVD to examine four global commodity futures markets — gold, wheat, WTI crude oil, and copper, demonstrating that the interconnectedness of commodities increases during financial crises, thereby weakening portfolio diversification.

The Granger causality tests are employed to determine whether past values of one stock's fundamentals can predict the future movements of another bank's stock price. This is essential for identifying leading indicators and potential spillover effects within our target sector. By establishing causality, we can better understand the mechanisms through which shocks propagate through the financial system, thereby enhancing our ability to predict and manage systemic risk. For instance, \citep{Cincinelli_Pellini_Urga_2022} used the Granger causality test to show that sharp price fluctuations in Chinese listed financial institutions can lead to spillover effects in the financial system, and severe declines in the Chinese stock market have become a major source of instability in global financial markets.

In the context of our study, the combination of FEVD and Granger causality tests was chosen to leverage their complementary strengths, allowing us to construct a comprehensive picture of the dynamic co-movements and causal relationships in high-frequency stock market data.

\subsection{The Forecast Error Variance Decomposition (FEVD) regressions}
For Finance research, FEVD is particularly useful in policy analysis, risk assessment, and economic forecasting. For instance, in a financial setting like bank risk control management, understanding how shocks to interest rates affect loan defaults, and vice versa, can be invaluable.

In the realm of Vector Autoregressive (VAR) modeling, Forecast Error Variance Decomposition (FEVD) is a commonly applied technique. A VAR(1) model can be expressed mathematically as:
\begin{equation}
\mathbf{Y}_t = \mathbf{A}_0 + \mathbf{A}_1 \mathbf{Y}_{t-1} + \mathbf{E}_t
\end{equation}
where \( \mathbf{Y}_t \) is a \( k \times 1 \) vector of endogenous variables, \( \mathbf{A}_0 \) is a constant term, \( \mathbf{A}_1 \) is a \( k \times k \) coefficient matrix, and \( \mathbf{E}_t \) is a \( k \times 1 \) vector of error terms.

The impulse response function captures the effect of a one-unit shock in one variable on all variables in the system at different time horizons. In the context of a VAR(1) model, the Impulse Response Function (IRF) can be directly computed from the coefficient matrix of the model. Specifically, for a VAR(1) model as described above, the IRF at time lag \( h \) is given by:

\begin{equation}
\text{IRF}(h) = \mathbf{A}_1^h
\end{equation}

Here, \( h \) represents the time lag and \( \mathbf{A}_1^h \) is the \( h \)-th power of the coefficient matrix \( \mathbf{A}_1 \). This function describes how the system would respond \( h \) periods after a unit shock is introduced in the error term \( \mathbf{E}_t \) at time \( t \).

The formula to compute FEVD usually involves decomposing the variance-covariance matrix of the forecast error into orthogonal components attributed to each shock. Often, the Cholesky decomposition is used for this purpose. Mathematically, this is represented as:
\begin{equation}
\text{FEVD}(h) = \frac{\sum_{i=1}^{k} \left( \text{IRF}_{ij}(h) \right)^2}{\text{Var}(\Delta Y_{j}(h))}
\end{equation}
\citep{Lanne_Nyberg_2016}

In the segment dedicated to Forecast Error Variance Decomposition (FEVD) analysis, we employ a nuanced approach to scrutinize the interconnectedness within the banking sector. Specifically, the dependent variable is constituted by the 5-minute excess return series for each individual bank. Conversely, the independent variables are composed of a weighted aggregate of 5-minute excess returns from a pool of all other banks. The weights are assigned based on the transaction volumes associated with each bank. To mitigate the influence of liquidity on stock price volatility, the output metrics are normalized by the respective transaction volumes.

The best lag order was automatically selected based on the Bayesian Information Criterion (BIC) during the modeling process.
In this paper, FEVD was analyzed over a 12-period (i.e., 1-hour) horizon. Our model was stable, meaning that the modulus of all the eigenvalues of the Vector Autoregression (VAR) model were within the unit circle.

The residuals showed no autocorrelation, and the model was stable, but the residuals did not meet the assumption of normal distribution.
The basic theory of VAR and FEVD does not require the data or residuals to conform to normal distribution. Therefore, theoretically, the non-normality of the residuals does not directly affect the FEVD results. This implies that FEVD can still provide useful information about the interactions among variables, even when residuals are not normally distributed.

The FEVD output divided by the logarithm of the average trading volume and the sum of FEVD were used as a set of dependent variables for our further research.

\subsection{A Comparative FEVD Analysis of Commercial, Regional, and State-owned Banks}
From FEVD analytical results, we have selected three representative banks for data visualization, based on criteria such as market share, geographic distribution, customer base, and financial performance. These banks are \textit{sh.600036} (Commercial Bank) \textbf{Table \ref{tab:stats_stability_sh600036}}, \textit{sz.002142} (Regional Bank) \textbf{Table \ref{tab:stats_stability_sz002142}}, and \textit{sh.601398} (State-owned Major Bank) \textbf{Table \ref{tab:stats_stability_sh601398}}. Collectively, they offer a comprehensive view of the dynamics within the FEVD sector. This selection highlights the diverse nature of the Chinese banking system, providing insights into the distinctive features of each bank type and underscores the interplay between different banking tiers within the Chinese financial landscape.

\subsubsection{Statistical and Model Stability Indicators: Comparative Analysis}

This section provides a consolidated review and comparison of the statistical indicators and model stabilities for three banks, reflecting their performance in time series forecasting.

Durbin-Watson Statistics:
For all three banks, the Durbin-Watson statistics approach the value of 2, indicating a lack of significant autocorrelation within their residuals. This suggests that each bank's model effectively captures the dynamics of its time series.

Jarque-Bera Test:
All three banks exhibit high Jarque-Bera statistics with negligible p-values, strongly rejecting the normality hypothesis, a common trait in financial time series due to their fat-tailed distributions.

Model Stability:
The eigenvalues of all three banks' VAR models are less than one in modulus, confirming the stability necessary for accurate economic interpretation and forecasting.

\begin{figure*}[htbp] 
  \centering 
  \includegraphics[width=\linewidth]{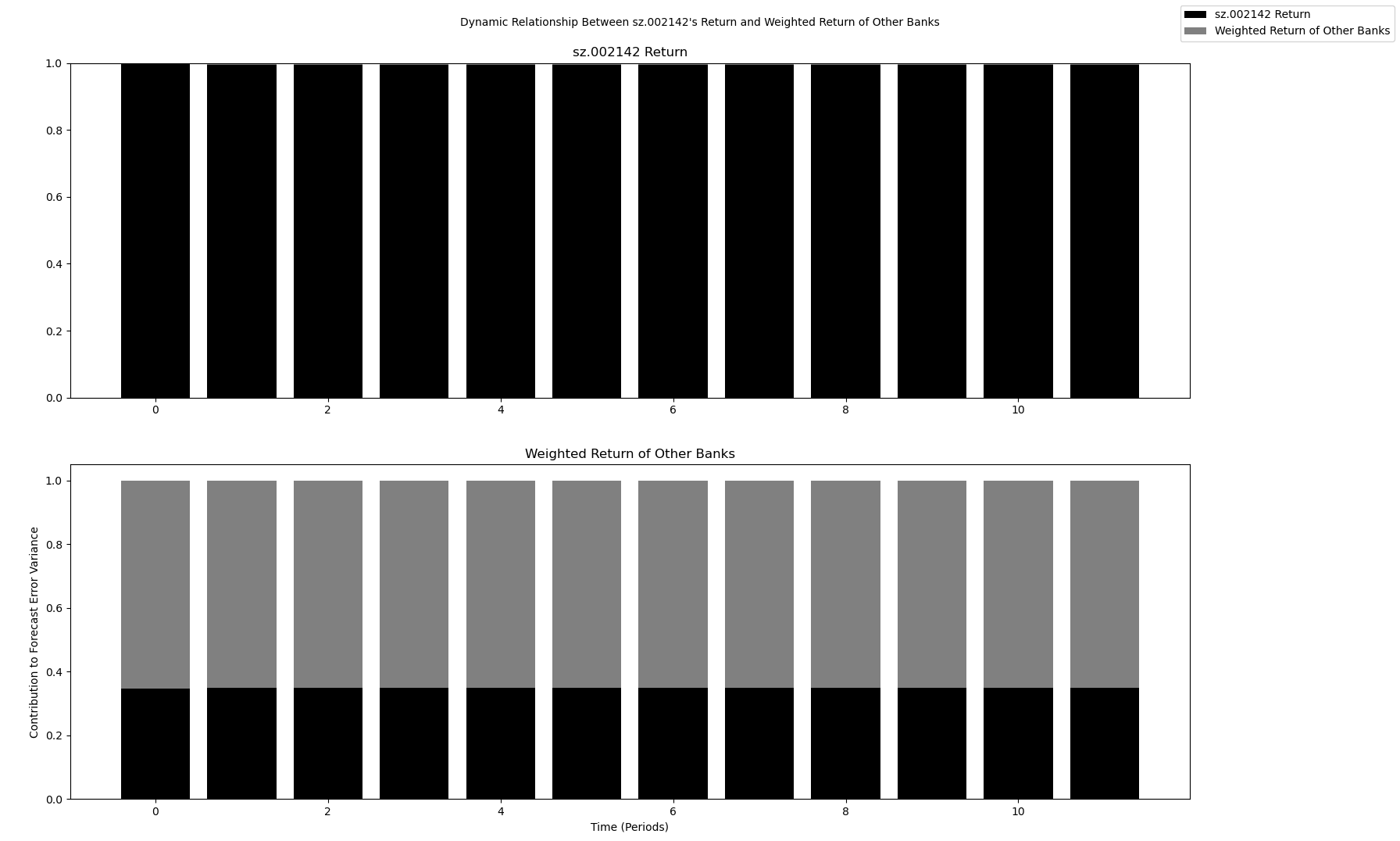} 
  \caption{FEVD Analysis Results for sz.002142.} 
  \label{fig:fevd_sz002142} 
\end{figure*}

\begin{figure*}[htbp] 
  \centering 
  \includegraphics[width=\linewidth]{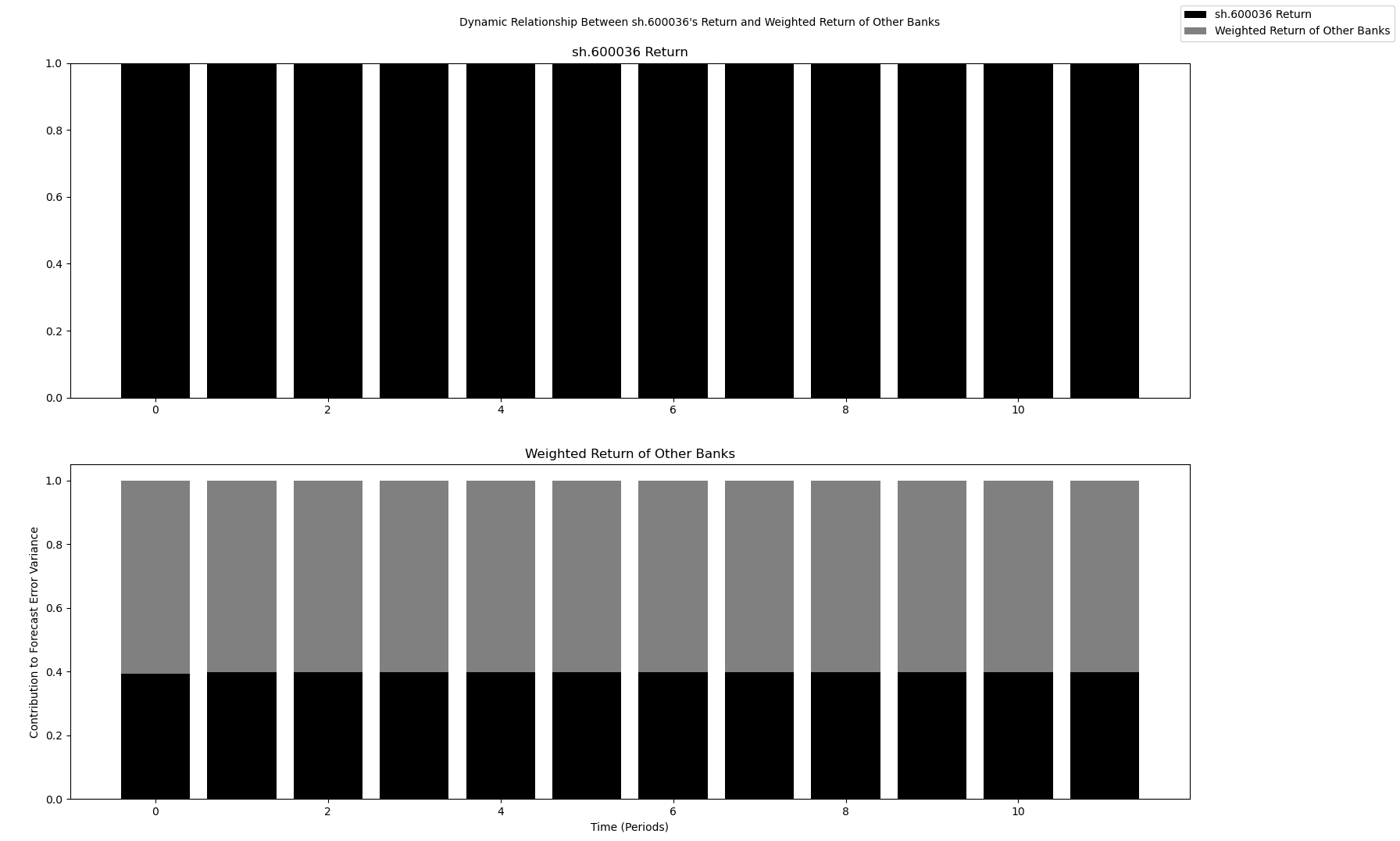} 
  \caption{FEVD Analysis Results for sh.600036.} 
  \label{fig:fevd_sh600036} 
\end{figure*}

\begin{figure*}[htbp] 
  \centering 
  \includegraphics[width=\linewidth]{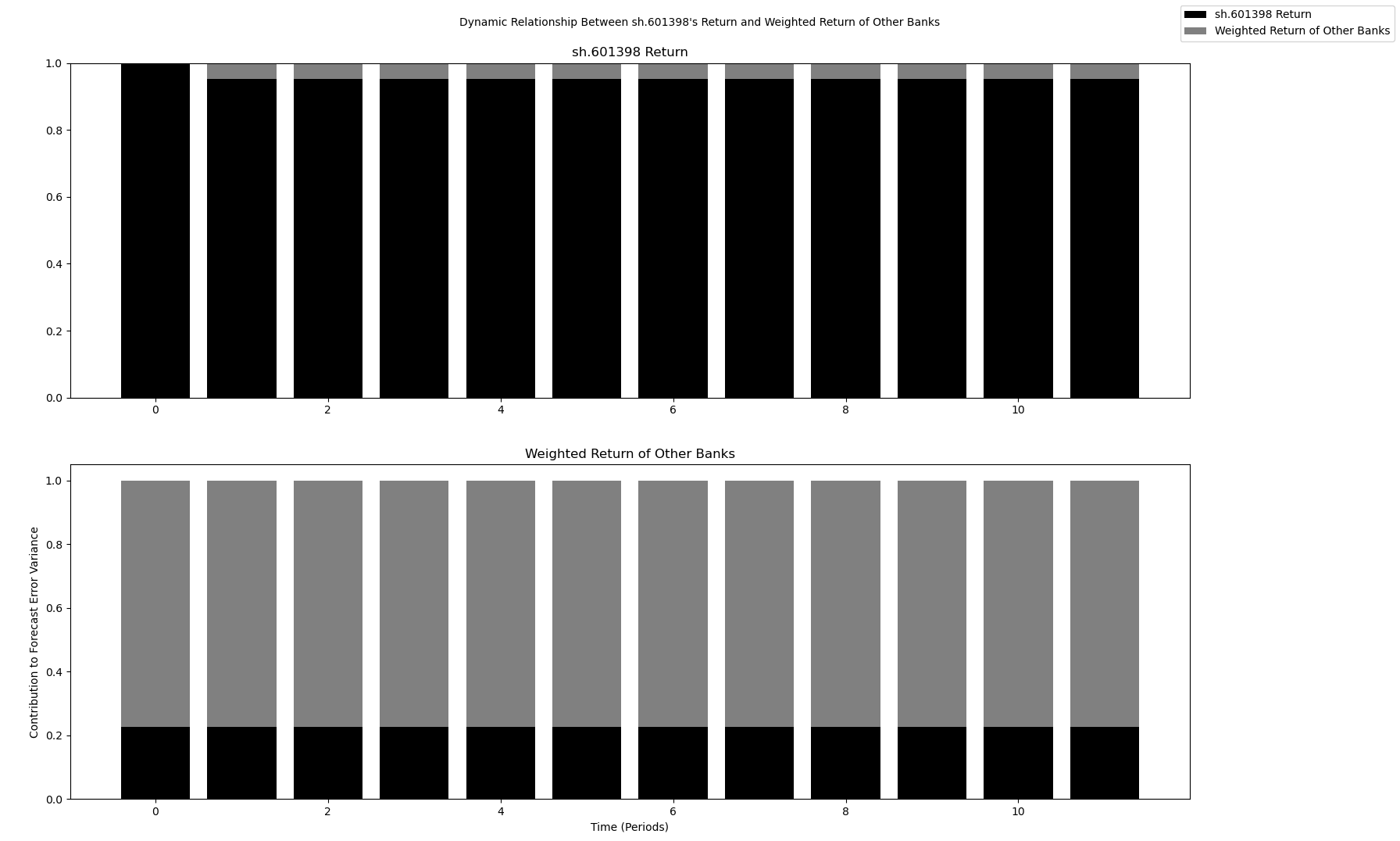} 
  \caption{FEVD Analysis Results for sh.601398.} 
  \label{fig:fevd_sh601398} 
\end{figure*}

Forecast Error Variance Decomposition (FEVD):
sz.002142 \textbf{Figure \ref{fig:fevd_sz002142}}: FEVD results demonstrate that its returns are mostly explained by its own past values, indicating minimal market-wide or sector-specific influences.
SH.601398 \textbf{Figure \ref{fig:fevd_sh601398}}: Initially, returns are solely explained by its own shocks, but over time, a slight external influence emerges, stabilizing around 95.15\% from its own past values.
sh.600036 \textbf{Figure \ref{fig:fevd_sh600036}}: The returns are largely self-explained (99.62\% stable), showing a very minimal impact from other banks.

Interbank Dynamics:
The influence of sz.002142 on other banks is minimal, yet the market dynamics exhibit a non-negligible impact on sz.002142, highlighting a degree of interconnectedness.
SH.601398 shows minor external influence, with about 22.66\% of its forecast error variance initially affected by other banks, stabilizing over time.
Conversely, sh.600036 impacts the weighted returns of other banks more significantly initially (39.24\%), but this stabilizes slightly higher at around 39.73\%.

Conclusion:
Overall, All of three models demonstrate no autocorrelation problems and are stable for further analysis. Each VAR model's residuals deviate from normality, typical in financial time series. FEVD analysis reveals that while individual banks are primarily influenced by their own historical shocks, there exists a varying degree of interconnectedness within the banking sector. sh.600036 shows the most significant impact on other banks, which may be attributed to its status as China's most successful business bank and its leading position in terms of trading volume and liquidity in the banking sector. Yet, all banks are predominantly driven by their own dynamics, with external influences remaining secondary.

\begin{table*}[htbp]
\centering
\caption{Summary of Statistics and Model Stability: sh.600036}
\label{tab:stats_stability_sh600036}
\begin{tabular}{lcc}
\hline
\textbf{Statistic/Test} & \textbf{Value/Result} \\
\hline
Durbin-Watson statistic & [1.9979, 2.0000] \\
Jarque-Bera (sh.600036 Return) & 997397.3546 \\
Jarque-Bera (Weighted Return) & 7307775.6395 \\
Model eigenvalues & [0.3600, 0.3107] & \\
Eigenvalues of VAR(1) rep & [0.0936, 0.0953] \\
\hline
\end{tabular}
\end{table*}

\begin{table*}[htbp]
\centering
\caption{Summary of Statistics and Model Stability: sh.601398}
\label{tab:stats_stability_sh601398}
\begin{tabular}{lcc}
\hline
\textbf{Statistic/Test} & \textbf{Value/Result} \\
\hline
Durbin-Watson statistic & [2.0023 2.0007] \\
Jarque-Bera (sh.601398 Return) & 975842.0098 \\
Jarque-Bera (Weighted Return) & 5673875.1298 \\
Model eigenvalues & [0.2639 0.1773] & \\
Eigenvalues of VAR(1) rep & [0.3469,
,0.3766
,0.3766
,0.3370
,0.3370
,0.3885] \\
\hline
\end{tabular}
\end{table*}

\begin{table*}[htbp]
\centering
\caption{Summary of Statistics and Model Stability: sz.002142}
\label{tab:stats_stability_sz002142}
\begin{tabular}{lcc}
\hline
\textbf{Statistic/Test} & \textbf{Value/Result} \\
\hline
Durbin-Watson statistic & [2.0010  1.9992] \\
Jarque-Bera (sz.002142 Return) & 471744.6355 \\
Jarque-Bera (Weighted Return) & 7442508.9832 \\
Model eigenvalues & [0.3728 0.3108] & \\
Eigenvalues of VAR(1) rep & [0.0965,
0.0980] \\
\hline
\end{tabular}
\end{table*}

\subsection{The Granger Causality Test}
In the section concerning Granger Causality tests, we adopt a methodological framework that aims to elucidate the causal relationships within the banking ecosystem. Specifically, the excess return for each bank serves as the dependent variable. As for the independent variables, we utilize the excess returns from other banks on the same trading day but lagged by one period, i.e., a 5-minute lag, to conduct the causality tests. 

For each lag order within a certain range (e.g., from 1 to 10), a Granger causality test was attempted. The Schwarz Information Criterion (SIC) was used to select the optimal lag order, i.e., the lag order that minimizes the SIC was selected as the optimal lag order. For each trading day, banks with the most causal impacts on other banks were counted and incremented by one to create a variable called \textquotedblleft significant001\textunderscore bool\textquotedblright. If there were multiple banks with the same highest number of influences on other banks on a single day, they were counted and incremented by one together. The total number of daily causal impacts on other banks throughout the year was logged as \textquotedblleft significant001\textunderscore times\textunderscore log\textquotedblright.

\subsection{Regression Analysis Steps}
One variable was selected from the dependent variable set as the dependent variable for each regression. The set of dependent variables, \( Y \), included:
\begin{enumerate}
  \item \texttt{"Significant001\textunderscore bool"} \textbf{Table~\ref{tab:table2}} 
  \item \texttt{"Significant001\textunderscore times\textunderscore log"} \textbf{Table~\ref{tab:table3}} 
  \item \texttt{"Sum FEVD Results"} \textbf{Table~\ref{tab:table4}} 
  \item \texttt{"Influence Per Unit Trade Amount"} \textbf{Table~\ref{tab:table5}} 
\end{enumerate}
The presence of multicollinearity was not pre-tested. Instead, all 192 independent variables were input at once during the regression. Variables were removed based on the condition number until it was below 100. In the process, all variables with a \( p \)-value above 0.5 were also excluded.

\section{Experimental Results}
\subsection{The number of days and the frequency (log-transformed) of interactions regression}
Significant001$\_$bool regression analysis in \textbf{Table~\ref{tab:table2}} concerning the total number of days in 2021 that each of all banks had a causal impact on others. 
Significant001$\_$times$\_$log regression analysis in \textbf{Table~\ref{tab:table3}} concerning the natural logarithm of the total number of times in 2021 that each of 37 banks had a causal impact on others. 

We will initially utilize a significant causal relationship analysis, marked at the 1\% significance level, focusing on the number of days and the frequency (log-transformed) of interactions among banks in 2021. This approach aims to uncover the intricate dynamics in bank stock price co-movements and their underlying correlations with fundamental banking data. The distinction between causality and correlation is crucial here, as our analysis seeks to unravel more than just surface-level associations. By examining both the frequency and the number of days of these interactions, we aim to gain insights into the market structure and interdependencies among financial institutions. The use of log transformation on frequency data aids in stabilizing variance, enhancing the robustness of our regression models. The choice of econometric models, including Boolean and log regression, is tailored to suit the nature of our data, ensuring effective and reliable conclusions. 

\begin{table*}[ht]
\centering
\caption{significant001$\_$bool Regression Results}
\label{tab:table2}
{\footnotesize
\begin{tabular}{lcr}
\toprule
\textbf{Variable} & \textbf{Coefficient} & \textbf{Significance} \\
\midrule
Constant & 27.6605 & *** \\
Proportion of Impairment Loss of Non-Credit Goods in Profits & -34.2440 & *** \\
Proportion of Asset Impairment Losses & 41.2697 & *** \\
Proportion of credit goods impairment losses in profits & 4.5611 & * \\
Wealth Management YoY & -18.6863 & ** \\
Wealth Management 21A vs 21H & 23.3091 & * \\
Proportion of Wealth Management in Total Assets & -2.2779 & ** \\
Interbank asset growth rate 21A compared to 21H & 5.9797 & * \\
Interbank Assets Growth Rate of 21A vs 20A & -3.7350 & * \\
Personal loans 21A ratio & 15.5145 &  \\
Real Estate & 13.4759 & *** \\
Demand Deposits in 2021 & -36.0139 & ** \\
Issued Bonds Interest & 8.3228 & *** \\
21A PE & -17.3739 & *** \\
Regulatory requirements for personal mortgages\_3 & -7.0732 & ** \\
Total loan regulatory requirements\_1 & 4.1501 & * \\
Decrease in Reserve Requirement Ratio Since December 2021\_1 & -4.9104 & * \\
Reduction in Reserve Ratio Boosts Interest Spread\_3 & -11.2561 & ** \\
Tier 1 Capital Regulatory Adequacy Ratio\_2 & 6.7148 & ** \\
Capital adequacy ratio regulatory requirement\_1 & 4.4757 & * \\
Years when core Tier 1 adequacy ratio hit regulatory floor\_1 & -12.6726 & ** \\
Year When Capital Adequacy Ratio Hits Regulatory Floor\_1 & 9.9328 & ** \\
Total Loan Still Needs to Be Reduced Ratio bool\_1 & 8.5300 & *** \\
\bottomrule
\textbf{Significance Levels:} * $p<0.1$; ** $p<0.05$; *** $p<0.01$

\end{tabular}
}
\end{table*}

\begin{table*}[ht]
\centering
\caption{significant001$\_$times$\_$log Regression Results}
\label{tab:table3}
{\footnotesize
\begin{tabular}{lcr}
\toprule
\textbf{Variable} & \textbf{Coefficient} & \textbf{Significance} \\
\midrule
Constant & 7.2881 & *** \\
Proportion of Impairment Loss of Non-Credit Goods in Profits & -1.1437 & *** \\
Proportion of Asset Impairment Losses & 1.1879 & ** \\
Proportion of credit goods impairment losses in profits & 0.1443 & * \\
Consumer Credit and Other Ratio & 0.7626 & * \\
Wealth Management YoY & -1.4713 & *** \\
Wealth Management 21A vs 21H & 1.3074 & *** \\
Proportion of Wealth Management in Total Assets & -0.1196 & *** \\
Interbank asset growth rate 21A compared to 21H & 0.2964 & *** \\
Interbank Assets Growth Rate of 21A vs 20A & -0.0470 &  \\
Personal loans 21A ratio & 0.3352 &  \\
Real Estate & 0.3663 & ** \\
Interbank Liabilities + Certificates Deposit in 2021 vs 20A & -0.4046 & *** \\
Demand Deposits in 2021 & -0.8604 & * \\
Issued Bonds Interest & 0.3027 & *** \\
21A PE & -0.7534 & *** \\
Total loan regulatory requirements\_1 & 0.2767 & *** \\
Decrease in Reserve Requirement Ratio Since December 2021\_1 & -0.1532 &  \\
Reduction in Reserve Ratio Boosts Interest Spread\_3 & -0.3613 & ** \\
Core Tier 1 adequacy ratio regulatory requirement\_4 & -0.3241 & ** \\
Tier 1 Capital Regulatory Adequacy Ratio\_2 & 0.1232 &  \\
Capital adequacy ratio regulatory requirement\_1 & 0.1054 &  \\
Years when core Tier 1 adequacy ratio hit regulatory floor\_1 & -0.3642 & ** \\
Years when Tier 1 capital adequacy ratio hit regulatory floor\_1 & -0.1205 &  \\
Years when Tier 1 capital adequacy ratio hit regulatory floor\_3 & -0.1311 &  \\
Year When Capital Adequacy Ratio Hits Regulatory Floor\_1 & 0.4692 & *** \\
Year When Capital Adequacy Ratio Hits Regulatory Floor\_2 & 0.0618 &  \\
Total Loan Still Needs to Be Reduced Ratio bool\_1 & 0.2398 & ** \\
Personal mortgage still needs to reduce ratio bool\_1 & 0.1141 &  \\
\bottomrule
\textbf{Significance Levels:} * $p<0.1$; ** $p<0.05$; *** $p<0.01$

\end{tabular}
}
\end{table*}

As we see the two results in \textbf{Table~\ref{tab:table2}} and \textbf{Table~\ref{tab:table3}}, through the categorization and analysis of variables with positive and negative coefficients, we can derive generalized conclusions about the characteristics of the two regression results.

\subsubsection{Common Patterns and Trends}
Risk Management and Market Influence: Across both models, variables related to risk management (such as the proportion of asset impairment losses and credit goods impairment losses) significantly impact a bank's market influence. Banks that are willing to undertake more risk in their core operations are often associated with increased influence in the banking sector.

Shift in Business Focus and Market Positioning: Variables that indicate a shift in business focus (like the growth in wealth management and increased proportion of non-credit goods impairment losses) are significant in both models. This suggests that banks overly concentrating on non-traditional operations might diminish their influence among other banks.

Regulatory Compliance and Market Trust: Variables related to regulatory compliance (such as the Tier 1 capital adequacy ratio and capital adequacy ratio regulatory requirements) are significant in both models. This reflects that adherence to regulatory requirements and maintaining healthy capital levels are crucial for a bank’s influence and trust in the market.

Growth Strategy and Competitive Positioning: Variables linked to growth strategies (such as real estate investments and bond interest income) are significant in both models. This demonstrates that banks can enhance their market influence through aggressive growth strategies, like expanding into real estate or increasing participation in the bond market.

\subsubsection{Consolidated Conclusion}
In conclusion, a bank's market influence is affected by a multitude of factors including risk management strategies, business focus, regulatory compliance, and growth strategies. A common theme in both models is that banks balancing risk and reward, adhering proactively to regulatory norms, and focusing on core business growth tend to have greater market influence. This reflects the necessity for banks in today's rapidly changing financial environment to find the right balance between staying competitive and ensuring robust operation.

\subsection{FEVD regressions}
While the Granger causality tests in the significant001$\_$bool and significant001$\_$times$\_$log models provide valuable insights into the temporal precedence and potential predictive power of various banking variables, they might not fully capture the intricate interdependencies and dynamic interactions inherent in financial data. Here, the VAR model's strength lies in its ability to model multiple time series simultaneously, offering a more holistic view of the system dynamics.

The FEVD aspect of the VAR model further enhances our analysis by decomposing the variance of each variable's forecast error into proportions attributable to shocks in every other variable in the model. This decomposition allows us to understand the relative importance of each variable in forecasting the others, over time.

We propose two specific implementations of the FEVD within a VAR framework:

\begin{enumerate}
    \item \textbf{Sum FEVD Regression for Bank Stock Prices (1-hour maximum lag)}: This model, referred to as \textbf{Table~\ref{tab:table4}}, will analyze the stock prices of all banks in 2021, capturing the immediate and lagged effects within a short-term horizon. This approach is particularly useful in understanding rapid market responses and the impact of short-term events.
    \item \textbf{Influence Per Unit Trade Amount Regression}: In this model, denoted as \textbf{Table~\ref{tab:table5}}, we take a novel approach by dividing each bank’s influence by the mean value of its transaction amount, using this as the dependent variable. This model offers a unique perspective, focusing on the relative influence of banks weighted by their transactional significance, thus providing insights into how transaction volumes might modulate market impacts.
\end{enumerate}

\begin{table*}[ht]
\centering
\caption{Sum FEVD Regression Results}
\label{tab:table4}
{\footnotesize
\begin{tabular}{lcr}
\toprule
\textbf{Variable} & \textbf{Coefficient} & \textbf{Significance} \\
\midrule
Constant & 0.7420 & ** \\
Corporate deposit ratio & -0.2502 &  \\
Proportion of Impairment Loss of Non-Credit Goods in Profits & 0.4496 & * \\
Proportion of Asset Impairment Losses & -0.1737 &  \\
Proportion of credit goods impairment losses in profits & -0.1464 & ** \\
Consumer Credit and Other Ratio & -0.1389 &  \\
Wealth Management 21A vs 21H & 0.7919 & ** \\
Proportion of Wealth Management in Total Assets & -0.0035 &  \\
Interbank asset growth rate 21A compared to 21H & 0.0790 &  \\
Personal loans 21A ratio & -0.4698 & * \\
Infrastructure category & 0.0984 &  \\
Real Estate & -0.0754 &  \\
Interbank Liabilities + Certificates Deposit in 2021 vs 20A & -0.0377 &  \\
Demand Deposits in 2021 & 0.5338 & * \\
Issued Bonds Interest & -0.1215 & * \\
Equity Multiplier Changes & 0.0109 &  \\
21A PE & -0.3083 & *** \\
Regulatory requirements for personal mortgages\_3 & 0.1254 &  \\
Total loan regulatory requirements\_1 & -0.1138 & * \\
Decrease in Reserve Requirement Ratio Since December 2021\_1 & -0.0402 &  \\
Reduction in Reserve Ratio Boosts Interest Spread\_3 & 0.2920 & * \\
Core Tier 1 adequacy ratio regulatory requirement\_4 & -0.1606 &  \\
Tier 1 Capital Regulatory Adequacy Ratio\_2 & 0.0361 &  \\
Years when core Tier 1 adequacy ratio hit regulatory floor\_1 & 0.2572 & * \\
Years when Tier 1 capital adequacy ratio hit regulatory floor\_1 & 0.0068 &  \\
Years when Tier 1 capital adequacy ratio hit regulatory floor\_3 & 0.3087 & ** \\
Year When Capital Adequacy Ratio Hits Regulatory Floor\_1 & -0.0779 &  \\
Year When Capital Adequacy Ratio Hits Regulatory Floor\_2 & -0.1063 &  \\
Total Loan Still Needs to Be Reduced Ratio bool\_1 & -0.3594 & *** \\
Personal mortgage still needs to reduce ratio bool\_1 & 0.0861 &  \\
\bottomrule
\textbf{Significance Levels:} * $p<0.1$; ** $p<0.05$; *** $p<0.01$
\end{tabular}
}
\end{table*}

\begin{table*}[ht]
\centering
\caption{Influence Per Unit Trade Amount Regression Results}
\label{tab:table5}
{\footnotesize
\begin{tabular}{lcr}
\toprule
\textbf{Variable} & \textbf{Coefficient} & \textbf{Significance} \\
\midrule
Constant & 0.0466 &  \\
ImpLossNonCredit & -0.0091 &  \\
CreditGoodsImpLoss & -0.0023 &  \\
WealthMgmtYoY & -0.0327 & ** \\
WealthMgmt21Avs21H & 0.0417 & ** \\
WealthMgmtInTotalAsset & -0.0031 &  \\
InfrastructCategory & 0.0028 &  \\
InterbankLiabPlusCD & -0.0182 & * \\
IssuedBondsInterest & 0.0095 &  \\
EquityMultipChange & 0.0037 &  \\
Bad/Delinquent & -0.0269 & ** \\
21A PE & -0.0078 & *** \\
TotalLoanRegReq\_1 & 0.0082 &  \\
TotalLoanRegReq\_3 & 0.0284 & ** \\
DecreaseInReserveReq\_1 & 0.0166 & * \\
ReducInReserveRatio\_1 & -0.0163 & * \\
ReducInReserveRatio\_2 & -0.0199 & * \\
CoreTier1\_1 & 0.0087 &  \\
CoreTier1\_2 & -0.0083 &  \\
Tier1Capital\_4 & -0.0296 & ** \\
CARRegReq\_3 & 0.0061 &  \\
YearsWhenCoreTier1\_1 & -0.0202 & * \\
YearsWhenTier1\_3 & 0.0193 & * \\
YearWhenCARF\_1 & 0.0242 & ** \\
PersonalMortgage\_1 & 0.0047 &  \\
YearsWhenCoreTier1\_1 & -0.0071 &  \\
\bottomrule
\textbf{Significance Levels:} * $p<0.1$; ** $p<0.05$; *** $p<0.01$
\end{tabular}
}
\end{table*}

\subsubsection{Common Conclusions from FEVD Models}
\textbf{Role of Risk and Financial Strategy}: Both models highlight the critical role of a bank's risk management strategies and financial operations (like issuance of bonds and handling of bad/delinquent loans) in influencing its stock market performance and broader market influence.

\textbf{Impact of Regulatory Compliance and Market Dynamics}: Variables related to regulatory requirements and market dynamics, such as reserve ratios and capital adequacy, are crucial in both models. They illustrate how compliance and strategic responses to market regulations can significantly impact a bank's influence.

\textbf{Significance of Asset Management and Growth Approaches}: The focus on wealth management and asset growth strategies, as indicated by variables related to wealth management and real estate investments, underscores the importance of these areas in shaping a bank's market presence and influence.

\textbf{Detailed Conclusion}: In synthesizing the findings from both models, it is evident that a bank's influence on stock prices and the financial market is a multifaceted phenomenon. It is shaped by a complex interplay of risk management, financial strategies, regulatory compliance, and growth approaches. Banks that adeptly manage their risk exposure, align their financial strategies with market and regulatory dynamics, and effectively balance their asset management and growth initiatives are likely to have a more significant influence on the stock market. This influence is further nuanced by the relative weight of their transactional activities, as seen in the "Influence Per Unit Trade Amount Regression" model.

\subsection{The results total summary}
From the four regression models, we can observe several general patterns. 

First, the growth of banks' wealth management business may reflect their strategies to expand their business scale and improve service quality, thereby exerting a certain influence on other banks. However, if the proportion of wealth management business in total assets is too high, it may lead to excessive reliance on a single business and lack of business diversification, thereby weakening its influence on other banks.

Second, the increase in interbank liabilities and interbank deposits may indicate the enhanced activity of banks in the interbank finance market. However, this could result in excessive reliance on interbank business, leading to a reduced influence on other banks.

The increase in the equity multiplier may reflect the level of banks' return on equity (ROE), thus potentially enhancing their influence on other banks. On the other hand, under the same conditions, banks with higher price-earnings ratios may have fewer causal relationships with other banks, indicating that bank investors undervalue banks with lower valuations and prioritize financial statement data.

The increase in non-performing/default loans may reflect a decline in banks' risk management capabilities, thus potentially reducing their influence on other banks.

The strictness of loan regulatory requirements may reflect regulatory agencies' focus on bank risks. Stricter loan regulatory requirements may prompt banks to enhance their risk management capabilities, thereby increasing their influence on other banks.

The decline in reserve requirement ratio may reflect improvements in the macroeconomic environment and relaxed regulatory policies. This may provide banks with more funds for business expansion, thereby increasing their influence on other banks. On the other hand, the reduction in reserve requirement ratio may lead to higher interest rate spreads, indicating improved profitability for banks. However, if this reduction is due to a decline in the reserve requirement ratio, it may reflect a decrease in banks' risk-bearing capacity, thereby potentially reducing their influence on other banks.

\section{Contributions and Practical Applications}

Our study distinguishes itself from the Fama-French research in several key aspects. While the Fama-French model primarily focuses on explaining long-term average stock returns across the entire market using risk factors such as market size and book-to-market ratios, our research zeroes in on the short-term comovement of stock prices within the banking sector. Specifically, we investigate whether the price fluctuations of certain bank stocks can predict or influence the price movements of other banks, with a particular emphasis on identifying the determinants that enable some bank stocks to emerge as "leader stocks." Additionally, our study uncovers banking-specific patterns and relationships, such as the impact of regulatory compliance rates and non-performing loan ratios on price leadership, which are not addressed by the Fama-French model. Furthermore, we incorporate the time dimension by examining the dynamic relationships of stock price movements over short-term periods, capturing transient market sentiments and industry-specific competitive dynamics that the Fama-French framework does not encompass.

The findings of our research offer several practical applications that extend beyond theoretical insights. Firstly, we propose the development of predictive models to identify potential leader stocks within the banking sector, enabling investors to allocate funds more effectively during market volatility to achieve excess returns. Secondly, based on the characteristics of leader stocks, we suggest designing more precise hedging strategies to mitigate portfolio risk by taking positions on leader and follower stocks according to different economic cycles. Additionally, our research can assist regulatory bodies in assessing the potential impact of new policies on the stock structure within the banking industry. Furthermore, the identification of leader stocks facilitates the creation of specialized banking indices, enhancing the accuracy of investment benchmarks and structured financial products. Lastly, our study provides a foundation for statistical arbitrage strategies by exploiting the correlated price movements between leader and follower stocks, allowing for risk-hedged trading opportunities.

\section{Conclusion}
In this study, we have developed an innovative framework to examine the interconnection between fundamental data and stock price co-movements, applying it to the Chinese banking sector's stock prices and fundamentals in the year 2021. The analysis revealed enlightening outcomes that contribute to our understanding of the complex dynamics within specific industry sectors. 

First, we introduce an innovative framework designed to systematically and scientifically detect stock price co-movement within the same industry. By leveraging extensive datasets and multiple models, we enhance the robustness of our analytical conclusions. We believe that our method significantly reduces the workload involved in analyzing co-movement of stock prices within the same industry. Moreover, it is highly portable and can be applied to any industry. We anticipate extending our research to explore stock price co-movement across different countries and industries in the near future.

Second, our research delineates the connections between a stock's fundamental characteristics and its capacity to influence other stocks. We demonstrate how intrinsic traits such as wealth management business, interbank activities, and equity multiplier significantly impact the co-movements within the banking sector \citep{barberisComovement2005}. Therefore, we shed light on how these intrinsic traits might affect the broader financial ecosystem. By employing high-frequency data \citep{engleDynamicConditionalCorrelation2002} and focusing on a cross-sectional perspective, this study departs from traditional daily or monthly time frames prevalent in the extant literature, providing a novel lens to view these phenomena \citep{acharyaCapitalShortfallNew2012}.

Third, the findings from our study provide pivotal insights for investment managers deliberating portfolio strategies. For instance, during bullish phases of the banking sector, our model suggests selecting bank stocks that exhibit the strongest ability to lead co-movement. Conversely, during bearish cycles, it recommends avoiding these leading stocks. By employing such a dynamic allocation approach, investment managers can substantially bolster portfolio performance—amplifying returns and reducing volatility—all while maintaining a consistent sectoral weight for banking\citep{10.1257/jep.23.1.77}.

Such strategic shifts in asset allocation, based on the cyclical dynamics of an industry, exemplify the nuanced strategies that can potentially optimize portfolio outcomes. Given the intricate interplay of factors in the financial markets, embracing such informed, data-driven strategies becomes paramount for investment success.

Nevertheless, we acknowledge the limitations of our study. Future research could incorporate panel data to generate a more robust and nuanced analysis of these relationships, which could further enrich our understanding of the forces shaping stock price co-movements. Moreover, in today's interconnected and complex financial world, a growing imperative exists to consider the potential impact of geopolitical risks on stock price co-movements within the banking sector \citep{Demirgüç-Kunt_Pedraza_Ruiz-Ortega_2021}. The ripple effects of international incidents—be it the trade disputes between China and the U.S., escalating geopolitical tensions worldwide, or the unexpected downfall of a major institution like Silicon Valley Bank \citep{Jiang_Matvos_Piskorski_Seru_2024,Aharon_Ali_Naved_2023}—can markedly sway co-movements in bank stocks \citep{bekaertRiskUncertaintyMonetary2013}. Grasping the implications of these events can extend beyond purely theoretical interest; they bear tangible, practical relevance for an international cohort of investors, financial managers, and policymakers \citep{pastorUncertaintyGovernmentPolicy2012}.

It is also crucial to recognize that the leading stock driving the comovement may vary depending on the cyclical state of the industry \citep{Choudhry_Papadimitriou_Shabi_2016}. In an upward industry cycle, stocks with promising growth prospects are likely to spearhead the industry's comovement. Conversely, in a downturn, stocks that are on the brink of bankruptcy or facing severe challenges could potentially lead the comovement. This dual-natured leadership in comovement necessitates a more versatile and dynamic framework for understanding and modeling these relationships \citep{Cieslak_Morse_Vissing-Jorgensen_2019}.

In conclusion, our study makes a substantive contribution to the literature on stock price co-movements, illuminating the intricate web of connections that bind banks and their fundamental characteristics. As we navigate an increasingly complex and interdependent financial landscape, such insights become even more critical. We hope that our findings spur further research in this area, catalyzing new approaches, theories, and applications for any stock sector, as long as focus on fundamental characteristics. More immediately, we trust that our work will empower investors with a sharper understanding and more effective tools for investment decisions and risk management within the banking sector, thereby contributing to the financial resilience and stability of our economies.

%
\begin{appendices}

\onecolumn  

\section{Summary Statistics of Independent Variables}
\begin{tiny} 
\begin{longtable}{|l|rrrrrrr|}
\caption{Summary Statistics of Independent Variables} \label{tab:A1} \\
\hline
\multicolumn{1}{|c|}{Variable name} & \multicolumn{1}{c|}{mean} & \multicolumn{1}{c|}{std} & \multicolumn{1}{c|}{min} & \multicolumn{1}{c|}{25\%} & \multicolumn{1}{c|}{50\%} & \multicolumn{1}{c|}{75\%} & \multicolumn{1}{c|}{max} \\ \hline
\endfirsthead

\multicolumn{8}{c}%
{{\bfseries \tablename\ \thetable{} -- continued from previous page}} \\
\hline
\multicolumn{1}{|c|}{Variable name} & \multicolumn{1}{c|}{mean} & \multicolumn{1}{c|}{std} & \multicolumn{1}{c|}{min} & \multicolumn{1}{c|}{25\%} & \multicolumn{1}{c|}{50\%} & \multicolumn{1}{c|}{75\%} & \multicolumn{1}{c|}{max} \\ \hline
\endhead

\hline \multicolumn{8}{|r|}{{Continued on next page}} \\ \hline
\endfoot

\hline \hline
\endlastfoot

\multicolumn{1}{|l|}{Past growth rate of corporate loans}                                      & \multicolumn{1}{r|}{0.12262162} & \multicolumn{1}{r|}{0.06153515} & \multicolumn{1}{r|}{0.012}      & \multicolumn{1}{r|}{0.086}      & \multicolumn{1}{r|}{0.112}      & \multicolumn{1}{r|}{0.153}      & \multicolumn{1}{r|}{0.296}      \\ \midrule
\multicolumn{1}{|l|}{Corporate Loan Proportion}                                                & \multicolumn{1}{r|}{0.56021622} & \multicolumn{1}{r|}{0.10385143} & \multicolumn{1}{r|}{0.326}      & \multicolumn{1}{r|}{0.503}      & \multicolumn{1}{r|}{0.559}      & \multicolumn{1}{r|}{0.631}      & \multicolumn{1}{r|}{0.805}      \\ \midrule
\multicolumn{1}{|l|}{Corporate loan business return rate}                                      & \multicolumn{1}{r|}{0.04664223} & \multicolumn{1}{r|}{0.00533063} & \multicolumn{1}{r|}{0.0384}     & \multicolumn{1}{r|}{0.0425}     & \multicolumn{1}{r|}{0.048}      & \multicolumn{1}{r|}{0.0493}     & \multicolumn{1}{r|}{0.0592}     \\ \midrule
\multicolumn{1}{|l|}{Comprehensive Cost Rate of Funds}                                         & \multicolumn{1}{r|}{0.02255676} & \multicolumn{1}{r|}{0.00310703} & \multicolumn{1}{r|}{0.0159}     & \multicolumn{1}{r|}{0.0217}     & \multicolumn{1}{r|}{0.0229}     & \multicolumn{1}{r|}{0.024}      & \multicolumn{1}{r|}{0.029}      \\ \midrule
\multicolumn{1}{|l|}{Corporate deposit ratio}                                                  & \multicolumn{1}{r|}{0.57975676} & \multicolumn{1}{r|}{0.17848738} & \multicolumn{1}{r|}{0.115}      & \multicolumn{1}{r|}{0.486}      & \multicolumn{1}{r|}{0.589}      & \multicolumn{1}{r|}{0.74}       & \multicolumn{1}{r|}{0.866}      \\
\bottomrule
\end{longtable}
\end{tiny} 

\section{Bank Fundamental Data One-Hot Variable Meaning Table}
{\small
\begin{longtable}{lll}
\caption{Bank Fundamental Data One-Hot Variable Meaning Table} \label{tab:bank_onehot_data} \\
\hline
\toprule
Variable & Value & One-Hot Value \\
\midrule
Regulatory requirements for personal mortgages & 32.5\% & 3 \\
Regulatory requirements for personal mortgages & 12.5\% & 2 \\
Regulatory requirements for personal mortgages & 20.0\% & 0 \\
Regulatory requirements for personal mortgages & 17.5\% & 1 \\
Total loan regulatory requirements & 40\% & 3 \\
Total loan regulatory requirements & 27.5\% & 0 \\
\bottomrule
\hline
\end{longtable}
}

\end{appendices}
%
%


\bibliographystyle{informs2014} 
\bibliography{sample.bib} 





\end{document}